# A simple and efficient explicit parallelization of logic programs using low-level threading primitives

Diptikalyan Saha

IBM Research Labs, Delhi, India
diptsaha@in.ibm.com

Paul Fodor

State University of New York at Stony Brook, USA
pfodor@cs.sunysb.edu

**Abstract**

In this work, we present an automatic way to parallelize logic programs for finding all the answers to queries using a transformation to low level threading primitives. Although much work has been done in parallelization of logic programming more than a decade ago (e.g., Aurora, Muse, YapOR), the current state of parallelizing logic programs is still very poor. This work presents a way for parallelism of tabled logic programs in XSB Prolog under the well founded semantics. An important contribution of this work relies in merging answer-tables from multiple children threads without incurring copying or full-sharing and synchronization of data-structures. The implementation of the parent-children shared answer-tables surpasses in efficiency all the other data-structures currently implemented for completion of answers in parallelization using multi-threading. The transformation and its lower-level answer merging predicates were implemented as an extension to the XSB system.

*Categories and Subject Descriptors* D.3.3 [**Programming Languages**]: Language Constructs and Features

*General Terms* Logic Programming.

## 1. Introduction

In the last few years, many applications are turning towards logic programming as a better way to represent knowledge and a logical semantics for huge amounts of data. Unfortunately, although logic programming works well for medium size programs, it is still lacking the capability to handle huge programs from areas like semantic web, data mining, program analysis, and security policy frameworks, so most researchers agree that we need to better understand parallelization of logic programs.

We believe that the area of logic programming was better studied more than a decade ago, when special WAM engines [3] were developed to run on multiple processors (see Aurora [4], Muse [1], and YapOR [5]). All these approaches extended the WAM in some way for shared memory architectures: Aurora kept a trail tree "derivation tree" and used "binding arrays" in each process to point to nodes in the trail tree, while Muse copied process bindings for each spawn thread. Unfortunately, during those times multiple core processors were extremely expensive and those approaches were never adopted by large communities or evolve into full-fledged systems. Lately, there is a higher interest in parallelization using common multi-core processors for answer-set programming. However, the work for well-founded models is still weak and we are aware of only one deductive database which allows parallelization of computation: OntoBroker. However, the improvements in efficiency are not very high (only about ~10% improvement for the parallel computation of the transitive closure of a very sparse random graph on a dual-core processor compared to a single process computation) due to its bottom-up computation and lower possibility for parallelization.

Multi-Core is one of the current buzz-words around along with related areas (e.g., Cloud Computing). It seems that the era of Moore's law is coming to an end and processor speeds are not going to increase as they way it was increasing in the pre-2002 era. Instead, the future performance gain is going to be exploiting the multiple CPUs available in the single processor by thread level parallelism. The main question here is the effect of this change in computing paradigm to the logic programming model. The WAM (Warren Abstract Machine) emulator and its tabled extensions such as SLG-WAM were designed primarily for a single threaded execution. The main problem is whether a tabled Prolog program can be compiled to a set of set of instructions where each set is going to run in a single CPU, and still preserve the soundness and completeness of tabled execution.

An important question to the programming language community is how to cope and exploit the power of parallel executions. In theory, we don't really need to write our programs along with parallel-programming constructs that the compiler can generate efficient code which exploits parallelism as the user knows what is the intension of the program, or the compiler take the burden of taking any program and generate efficient parallel code from it, after all the compiler knows more about the system and architecture. We believe that it's a new opportunity for declarative community where programmer, by definition, writes down its intent through a declarative language and then the compiler can take the burden of generating efficient code which exploits parallelism in the executing machine.

Many LP systems support the use of POSIX threads to perform separable computations with a simple and clear API (e.g., Yap, SWI, XSB, and Mercury). Our work describes parallelization of computing tabled predicates using a transformation into a parallel version which distributes the computation on multiple threads. This transformation for multi-core processors can be extended to a multi-computer MapReduce scheme [2] (i.e.,"map" for distributing the data and "reduce" to combine the results). Due to its simple join for the "reduce" operation, such a transformation has a

## 2. Parallelization through explicit threading and answer-tries merging

In this section we describe the parallelization of computing models for tabled predicates. Our work was inspired from the map-reduce algorithm, although not a new paradigm, it is a simple and efficient way to parallelize computations. Intuitively, our method can be described as follows: we split the query data and distribute the data to multiple computers/threads (in map-reduce terms, this can be viewed as the "map" phase), keep the results sets disjoined and join and combine the results (i.e., the "reduce" phase in map-reduce terms).

The motivating example for this work is the right recursive transitive closure and we consider the simplest query to this program: reach(bound,Y) with only the first argument bound. In most cases this is the most used by programmers. This program is easily transformed into a parallel multi-threaded version as seen in Program 2 bellow.

```
:- table(reachr/2).
reachr(X,Y):- edge(X,Y).
reachr(X,Y):- edge(X,Z), reachr(Z,Y).
```
**Program 1: Right recursive transitive closure**

```
:- table(reachr/2).
reachr(X,Y):-
  thread_create((
    edge1(X,Z1),
    reachr1(Z1,_),
    merge_ans_tables(reachr(X,_), reachr1(Z1,_))),Id1),
  thread_create((
    edge2(X,Z2),
    reachr2(Z2,_),
    merge_ans_tables(reachr1(X,_), reachr2(Z2,_))),Id2),
  ...
  thread_join(Id1),
  thread_join(Id2),
  ...
  reachr(X,Y). % tabling suspension to collect the results

:- table reachr1/2.
reachr1(X,Y):- edge(X,Y).
reachr1(X,Y):-
  edge(X,Z),
  check_if_not_tried(reachr(X,Z)),
  reachr1(Z,Y).
...
:- thread_shared_constant edge/2.
:- import check_if_not_tried/1, merge_ans_tables/3 from increval.
```
**Program 2: Right recursive tc parallel transformation**

The first step of our algorithm is the initial distribution of tasks to the children threads. Considering the program 1, the transformation looks for a distribution of the computation in the first step, so it looks for two or more edges starting from the initial bound node in the graph. If such a case is not found, the algorithm will continue to compute the transitive closure and keep looking for 2 or more edges from the same node before putting those options into the children threads.

The second step is to run the children threads in parallel on the complete data set (i.e., all edges). However, one essential condition for efficiency is that computations do not overlap (i.e., without duplication their tasks), and their answer-tables have no results in common and making sure that if one thread stops then it takes up other thread's work.. This is done building the Predicate Dependency Graph (PDL) and adding in front of each literal $L$ that participates in a cycle in PDL a safe version of a semaphore "*check_if_not_tried(L)*" which checks if the query was not already started by another child thread. If it was not started, it is marked as started and the program proceeds, otherwise the computation of the thread fails and the computation backtracks. This incurs that the computation for each thread is sound, but the result set is not complete. However, the union of all the answer sets of the children threads is complete.

The third step is the merging of the answers from all the children threads (for completion). After all the children threads have completed their respective answer tables, we merge the answer tables to return the answers in the parent threads table. This is a very common operation for programs that find all answers to queries or compute various aggregate functions.

Using full-shared data-structures (i.e., shared tries in XSB) is impractical because it induces a really significant overhead for copying results from all threads. This overhead is over 30% for copying a single tabled answer-table for the transitive closure example in XSB.

We implemented a method that merges answer tables in constant time. On completion of the query to the parent thread (i.e., after the join with all its children and end of querying), we remove the trie based return of the answers. In a builtin-call each thread's answer table *ponter_next* points to the other threads table answer

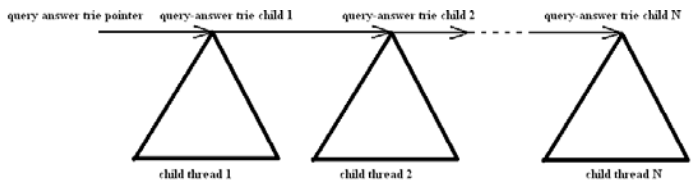

tables *head_pointer*. Finally, we return answers through the answer table pointer when asked for answers in the parent table.

## 3. Evaluation Results

We evaluated the performance of the parallelism method for transitive closure programs and for real examples from program analysis with the Andersen's Points-to Analysis which uses two flavors of transitive closure: left and right recursive.

First, we compared the performance of transitive closure for single process computation (i.e., un-threaded) for XSB, parallelized multi-threading computation without lower-level answer-tables merging for XSB (i.e., using shared predicates and, basically, copying of answers), parallelized computation with multi-threading with lower-level answer-tables merging for XSB. On a dual-core Dell Optiplex 755 with 3GB main memory running Ubuntu 9.10, we computed transitive closures of complete graphs for sizes between 100 and 10,000 vertexes, obtaining an improvement of 40.51% in average for the wall time with two threads. For instance, for 10K vertexes, we got 4.89sec. for the computation on two threads vs. 8.22 for the single-threaded computation.

The points-to analysis experiment shows an even better improvement (see table 1 for the data set *Vasy59* data set).

|  |  |  | **Total time** |
|---|---|---|---|
| Single-threaded computation | | | 2.28 sec |
| 2 threads with shared predicates for merging results | | | 1.64 sec |
| 2 threads | Thread1 | 0.83 | 1.19 sec |
| | Thread2 | 1.05 | |
| | Merging | 0.14 | |

*Table 1: Execution times for points-to analysis*

## 4. Related Work

The work by Marques and Swift in [8] is the closest research to our own work and it's the basis for our implementation in XSB Prolog. However, their work doesn't actually address parallelization of tabled computations, but instead, correctness and completeness of SLG resolution for shared tabled predicate calls in multiple threads. Their algorithms deal with isolating the computation space for repetitive computations and they show how repetitive calls to the same tabled calls are "usurped" (i.e., stopped) until the first thread to call the query completes the computation. However, our implementation also deals with merging of answer tables, which is a way to "promote" answers computed by all children threads and adjusting the space reclamation routines to account for space that is "promoted", done concurrently by all children threads. An alternative to promoting answers is to use private message queues, but we are confident that our method has better efficiency due to its constant cost.

[6] and [7] share the same basic ideas with [8], that is multiple similar tabled calls are computed in a single thread. This is again different from our approach because we can still distribute different parts of the computation for the specific case of transitive closure.

Other approaches to parallelism of logic programs (such as: [1,2,4,5] and the works cited in [3]) are just distantly related to our approach, although the main goal was the same with ours. The main difference is that all these approaches modify the Warren Abstract Machine (copying environments, binding arrays, back tracing, etc.) while we use explicit transformation with multi-threading.

## 5. Conclusions and Future Work

We presented a novel approach that transforms logic programs under the well founded semantics into parallelized versions and has been implemented in the increval package of the XSB Prolog system (available for download from [10]). The primary advantages of this approach are its good efficiency and clear transformation while preserving the soundness and completeness.